\documentstyle[twoside,fleqn,espcrc2,graphicx]{article}

\title{A new treatment of the in-medium chiral condensates}

\author{G.~X.~Peng$^{\mathrm{a,b}}$,
        M.~Loewe$^{\mathrm{c}}$,
        U.~Lombardo$^{\mathrm{d}}$,
        X.~J.~Wen$^{\mathrm{b}}$   \\[0.25cm]
    $^{\mathrm{a}}$
 China Center of Advanced Science and Technology
 (World Laboratory), Beijing 100080, China\\
    $^{\mathrm{b}}$
 Institute of High Energy Physics,
 Chinese Academy of Sciences, Beijing 100039, China\\
     $^{\mathrm{c}}$
 Facultad de Fisica, Pontificia Universidad Cat\'{o}lica
 de Chile, Casilla 306, Santiago 22, Chile\\
     $^{\mathrm{d}}$
 Dipartimento di Fisica, 57 Corso Italia,
 and INFN-LNS, Via Santa Sofia, 9500 Catania, Italy
        }

\begin{document}

\begin{abstract}
A new formalism to calculate the in-medium chiral condensate
is presented. At lower densities, this approach leads to a linear
expression. If we demand a compatibility with the famous model-independent
result, then the pion-nucleon sigma term should be six times the average
current mass of light quarks. QCD-like interactions may slow the
decreasing behavior of the condensate with increasing densities,
compared with the linear extrapolation, if densities are lower than
twice the nuclear saturation density. At higher
densities, the condensate vanishes inevitably.
\end{abstract}

\maketitle

The behavior of chiral condensates in a medium has
been an interesting topic in nuclear physics \cite{review}.
 A popular method to calculate
the in-medium quark condensate is the Feynman-Helmann theorem.
The main difficulty, however, is the assumptions we have to make
on the derivatives of model parameters with respect to the quark
current mass.

To bypass this difficulty, we will apply a similar idea as in
the study of strange quark matter \cite{Peng,Surk,Ben,Cha}
by defining an equivalent mass. A differential equation which
determines the equivalent mass will be derived.
At lower densities, the new formalism leads to a linear decreasing
condensate. A comparison with the result in nuclear matter
implies that the pion-nucleon sigma term should be six times the
average current mass of light quarks. At higher densities,
it turns out that the decreasing speed of the condensate with
increasing densities is lowered, compared with the linear extrapolation.

 The QCD Hamiltonian density can be schematically written as
\begin{equation}     \label{Hqcd}
 H_{\mathrm{QCD}}
= H_{\mathrm{k}} + \sum_i m_{i0}\bar{q_i}q_i + H_{\mathrm{I}},
\end{equation}
where $H_{\mathrm{k}}$ is the kinetic term,
$m_{i0}$ is the current mass of quark flavor $i$,
and $H_{\mathrm{I}}$ is the interacting part of the Hamiltonian.
The sum goes over all flavors involved.

The basic idea of the mass-density-dependent model of quark
matter is that the system energy can be expressed as in
a noninteracting system, where the strong interaction implies a
variation of the quark masses with density.
In order not to confuse with other mass concepts, let us call such a
density-dependent mass as an equivalent mass.
It can be separated into two parts, i.e.,
\begin{equation}
m_i=m_{i0}+m_{\mathrm{I}},
\end{equation}
where the first term is the quark current mass and the second part
is a flavor independent interacting part.
Therefore, we will have
a Hamiltonian density of the form
\begin{equation}    \label{Hequiv}
  H_{\mathrm{eqv}}=H_{\mathrm{k}} + \sum_i m_i \bar{q_i}q_i.
\end{equation}

We require that the two Hamiltonian densities
$H_{\mathrm{eqv}}$ and $H_{\mathrm{QCD}}$ should have the same
expectation value for any state $|\Psi\rangle$, i.e.,
\begin{equation}
  \langle{\Psi} | H_{\mathrm{eqv}} |\Psi\rangle
 =\langle{\Psi} | H_{\mathrm{QCD}} |\Psi\rangle.
\end{equation}
Applying this equality to the state $|n_{\mathrm{B}}\rangle$\
with baryon number density $n_{\mathrm{B}}$
and to the vacuum state $|0\rangle$,
we have
\begin{equation} \label{Hrel}
  \langle H_{\mathrm{eqv}}\rangle_{n_{\mathrm{B}}}
 -\langle H_{\mathrm{eqv}}\rangle_0
 =\langle H_{\mathrm{QCD}}\rangle_{n_{\mathrm{B}}}
 -\langle H_{\mathrm{QCD}}\rangle_0.
\end{equation}
Here  we use $
\langle A\rangle_{n_{\mathrm{B}}}
\! \! \equiv \! \! \langle n_{\mathrm{B}}|A|n_{\mathrm{B}}\rangle
$
and $
\langle A\rangle_0
\! \!\equiv \! \! \langle 0|A|0\rangle
$ for an arbitrary operator $A$.

   We restrict  ourselves to systems with uniformly distributed
particles where we can write
$
\langle \Psi|m(n_{\mathrm{B}})\bar{q}q|\Psi\rangle
=m(n_{\mathrm{B}})\langle \Psi|\bar{q}q|\Psi\rangle.
$
Accordingly we can solve Eq.\ (\ref{Hrel}) for $m_{\mathrm{I}}$,
getting
\begin{equation} \label{mIdef}
m_{\mathrm{I}}
=\frac{\epsilon_{\mathrm{I}}}
      {\sum_i (\langle\bar{q_i}q_i\rangle_{n_{\mathrm{B}}}
       -\langle\bar{q_i}q_i\rangle_0)},
\end{equation}
where $\epsilon_{\mathrm{I}}\equiv
 \langle H_{\mathrm{I}}\rangle_{n_{\mathrm{B}}}
  -\langle H_{\mathrm{I}}\rangle_0$
is the interacting energy density.

Therefore, considering the quarks as a free system, while keeping the
system energy unchanged, they should acquire an
equivalent mass corresponding to the current mass plus the common
interacting part shown in Eq.\ (\ref{mIdef}).
The equivalent mass is a function of
the quark current mass and of the density.
Note that quark confinement
implies the following natural requirement:
\begin{equation} \label{mlim}
\lim_{n_{\mathrm{B}}\rightarrow 0} m_{\mathrm{I}}=\infty.
\end{equation}

Because the Hamiltonian density $H_{\mathrm{eqv}}$ has the form
of a system of free particles with equivalent masses $m_i$,
the energy density of quark matter can be expressed as
\begin{eqnarray}   \label{epsilon}
 \epsilon
&=&\sum_i \frac{g}{2\pi^2} \int^{k_{\mathrm{f}}}_0
         \sqrt{k^2+m_i^2}\ k^2 dk \nonumber\\
&=& 3n_{\mathrm{B}} \sum_i m_i
   F\left(\frac{k_{\mathrm{f}}}{m_i}\right), \label{epsi}
\end{eqnarray}
where $g=3$(colors)$\times$2(spins) = 6 is the
degeneracy factor, and
\begin{equation} \label{kfdef}
k_{\mathrm{f}}=\left(\frac{18}{g}\pi^2n_{\mathrm{B}}\right)^{1/3}
\end{equation}
is the Fermi momentum of the quarks. The function $F(x)$ is
\begin{equation} \label{Fdef}
F(x) \equiv \frac{3}{8}[x\sqrt{x^2+1}\,(2x^2+1)
- \mbox{sh}^{-1}(x) ]/x^{3}.
\end{equation}
For convenience, let us define the function $f(x)$
\begin{eqnarray} \label{fdef}
f(x)&\equiv& -x^2d[F(x)/x]/dx \nonumber\\
    &=& \frac{3}{2}\left[
        x\sqrt{x^2+1}-\mbox{sh}^{-1}(x)
                   \right]/x^3.
\end{eqnarray}

On the other hand, the total energy density can also be
expressed as
\begin{eqnarray}
\epsilon &=& \sum_i \frac{g}{2\pi^2}
   \int_0^{k_{\mathrm{f}0}}\sqrt{k^2+m_{i0}^2}k^2 dk
           +\epsilon_{\mathrm{I}}  \ \nonumber\\
  &=& \frac{3}{N_{\mathrm{f}}}n_{\mathrm{B}}\sum_i m_{i0}
       F\left(\frac{k_{\mathrm{f}0}}{m_{i0}}\right)
       +\epsilon_{\mathrm{I}},  \label{ep0}
\end{eqnarray}

where $N_{\mathrm{f}}=2$ is the number of flavors,
the first term is the energy density without
interactions, and the second term is the interacting part.
Because $n_{\mathrm{B}}/N_{\mathrm{f}}$ is the baryon number density
for each flavor, the Fermi momentum $k_{\mathrm{f}0}$ here is
\begin{equation}
k_{\mathrm{f}0}
=\left(\frac{18}{g} \pi^2
 \frac{n_{\mathrm{B}}}{N_{\mathrm{f}}}\right)^{1/3}.
\end{equation}
It looks similar to the non-interacting
system. However, the Fermi momentum $k_{\mathrm{f}}$ in
Eq.\ (\ref{kfdef}) is bigger. It has been boosted
because of the Fermi momentum dependence
on density through the equivalent mass. In the appendix A,
we will give a proof for the boosting of the Fermi momentum.

Combining Eqs.\ (\ref{ep0}) and (\ref{epsilon}) we identify
\begin{equation} \label{eImI}
\frac{\epsilon_{\mathrm{I}}}{3n_{\mathrm{B}}}
=\sum_i \left[
   m_i F\left(\frac{k_{\mathrm{f}}}{m_i}\right)
   - \frac{m_{i0}}{N_{\mathrm{f}}}
     F\left(\frac{k_{\mathrm{f}0}}{m_{i0}}\right)
        \right].
\end{equation}

The Hellmann-Feynman theorem gives
\begin{equation} \label{helfey}
 \langle\Psi|\frac{\partial}{\partial\lambda}
  H(\lambda)|\Psi\rangle
  =\frac{\partial}{\partial\lambda} \langle\Psi|
   H(\lambda)|\Psi\rangle,
\end{equation}
where $|\Psi\rangle$\ is a normalized eigenvector of the Hamiltonian
 $H(\lambda)$ which depends on a parameter $\lambda$.

  On application, in Eq.\ (\ref{helfey}), of the substitutions
$\lambda\rightarrow m_{i0}$
and $H(\lambda) \rightarrow \int d^3x H_{\mathrm{QCD}}$,
one gets
$ 
 \langle\Psi| \int d^3x \bar{q_i}q_i|\Psi\rangle
  =\frac{\partial}{\partial m_{i0}} \langle\Psi|
   \int d^3x H_{\mathrm{QCD}} |\Psi\rangle
$ 
for each flavor $i$.
   Applying this equality, respectively, to the
state  $|n_{\mathrm{B}}\rangle$ (quark matter with baryon number
density $n_{\mathrm{B}}$) and to the vacuum $|0\rangle$, one obtains
\begin{equation}  \label{qcCohen}
 \langle\bar{q_i}q_i\rangle_{n_{\mathrm{B}}}
 -\langle\bar{q_i}q_i\rangle_0
  = \frac{\partial\epsilon}{\partial m_{i0}},
\end{equation}
where
$
\epsilon
\equiv\langle H_{\mathrm{QCD}}\rangle_{n_{\mathrm{B}}}
 -\langle H_{\mathrm{QCD}}\rangle_0
$
is the total energy density.
Now let us substitute Eq.\ (\ref{epsilon})
into Eq.\ (\ref{qcCohen}), carry out the
corresponding derivative, and sum over the flavor index.  We get
\begin{eqnarray}
&& \sum_i \left[
 \langle\bar{q_i}q_i\rangle_{n_{\mathrm{B}}}
 -\langle\bar{q_i}q_i\rangle_0
       \right]      \nonumber\\
&&\phantom{abc}
=3n_{\mathrm{B}}\sum_i f
  \left(\frac{k_{\mathrm{f}}}{m_i}\right)
  \left[1+\nabla m_{\mathrm{I}}
  \right]
\label{qcsum1}
\end{eqnarray}
with $\nabla\equiv \sum_i \partial/\partial m_{i0}$.
Note that $\nabla$\ is a differential operator
in mass space.


Comparing this equation with Eq.\ (\ref{mIdef}) we have
\begin{eqnarray} \label{delmIeI}
\nabla m_{\mathrm{I}}
=\frac{\epsilon_{\mathrm{I}}/(3n_{\mathrm{B}})}
      {m_{\mathrm{I}}\sum_i
        f\left(k_{\mathrm{f}}/m_i\right)}
 -1.
\end{eqnarray}

Replacing $\epsilon_{\mathrm{I}}/(3n_{\mathrm{B}}$ here by the right
hand side of Eq.\ (\ref{eImI}) we get
a first order differential equation
for the interacting equivalent mass.

Such a mass really exists, and we can
prove that it can be expressed in terms of the interacting
energy density $\epsilon_{\mathrm{I}}$ formally as
\begin{equation} \label{mIeI}
m_{\mathrm{I}}=\frac{\epsilon_{\mathrm{I}}/(3n_{\mathrm{B}})}
                    {
\frac{\nabla\epsilon_{\mathrm{I}}}{3n_{\mathrm{B}}}
+\frac{1}{N_{\mathrm{f}}}
 \sum_if\left(\frac{k_{\mathrm{f}0}}{m_{i0}}\right)
                    }.
\end{equation}

In the flavor symmetric case, i.e.,
$m_{u0}=m_{d0}=\cdots=m_0$, we have
$m_u=m_d=\cdots=m$, $\langle\bar{q_u}q_u\rangle
=\langle\bar{q_d}q_d\rangle
=\cdots
=\langle\bar{q}q\rangle$,
and $\nabla=\partial/\partial m_0 $. In this case
\begin{equation}  \label{mIeq2}
\frac{\partial m_{\mathrm{I}}}{\partial m_0}
=\frac{ m F\left(k_{\mathrm{f}}/m\right)
   -\frac{m_0}{N_{\mathrm{f}}}
    F\left(k_{\mathrm{f}0}/m_0\right) }
      {m_{\mathrm{I}}
       f\left(k_{\mathrm{f}}/m\right)}
 -1,
\end{equation}
\begin{equation}  \label{qcmI2}
\frac{\langle\bar{q}q\rangle_{n_{\mathrm{B}}}}
     {\langle\bar{q}q\rangle_0}
=1+\frac{1}{N_{\mathrm{f}}\langle\bar{q}q\rangle_0}
  \frac{\epsilon_{\mathrm{I}}}{m_{\mathrm{I}}},
\end{equation}
\begin{equation} \label{eImI2}
 m F\left(\frac{k_{\mathrm{f}}}{m}\right)
  -\frac{m_0}{N_{\mathrm{f}}}
   F\left(\frac{k_{\mathrm{f}0}}{m_0}\right)
=\frac{\epsilon_{\mathrm{I}}}{3N_{\mathrm{f}}n_{\mathrm{B}}}.
\end{equation}

Since $\lim_{x\rightarrow 0}F(x)=1$,
Eq.\ (\ref{eImI2}) becomes at lower densities
\begin{equation} \label{mlowden}
m=m_0+
\frac{\epsilon_{\mathrm{I}}}{3N_{\mathrm{f}}n_{\mathrm{B}}}.
\end{equation}
This means
$
m_{\mathrm{I}}
=\epsilon_{\mathrm{I}}/(3N_{\mathrm{f}}n_{\mathrm{B}})
$,
i.e.,
$
\epsilon_{\mathrm{I}}/m_{\mathrm{I}}
=3N_{\mathrm{f}}n_{\mathrm{B}}.
$
Substituting this ratio into Eq.\ (\ref{qcmI2}), we get
\begin{equation}  \label{qcL}
\frac{\langle\bar{q}q\rangle_{n_{\mathrm{B}}}}
     {\langle\bar{q}q\rangle_0}
=1-\frac{n_{\mathrm{B}}}{n^*}
\end{equation}
with
\begin{equation} \label{nstar}
n^* \equiv -\frac{1}{3}\langle\bar{q}q\rangle_0
=\frac{m_{\pi}^2f_{\pi}^2}{6m_0},
\end{equation}
where $m_{\pi}\approx 140$ MeV is the pion mass and
$f_{\pi}\approx 93.2$ MeV is the pion decay constant.

Since we have said nothing about the form of the interacting
energy density, our result is model independent. Recalling
that there is a model-dependent result in nuclear matter, i.e.,
\begin{equation} \label{qcnlin}
\frac{\langle\bar{q}q\rangle_{\rho}}{\langle\bar{q}q\rangle_0}
=1-\frac{\rho}{\rho^*}
 \ \mbox{with} \
\rho^*\equiv\frac{M_{\pi}^2F_{\pi}^2}{\sigma_{\mathrm{N}}},
\end{equation}
first proposed by Drukarev {\sl et al.}\
 \cite{Drukarev99}, and later re-justified by many authors
\cite{Cohen91},
we get, from the requirement $n^*=\rho^*$,
the very interesting relation
$\sigma_{\mathrm{N}}=6m_0$,
i.e., the pion-nucleon sigma term $\sigma_{\mathrm{N}}$
is six times the average current quark mass $m_0$.
If one takes $\sigma_{\mathrm{N}}=45$ MeV
\cite{Gasser91,Gensini80,Gibbs98}
and $m_0=(m_{u0}+m_{d0})/2=(5+10)/2=7.5$ MeV \cite{Gasser82},
we confirm this result.

The chiral condensate
at higher densities can be calculated from
Eqs.\ (\ref{mIeq2})-(\ref{eImI2})
if we know the interacting energy density
$\epsilon_{\mathrm{I}}$ from a realistic quark model.
In the following, we consider a simple example.

Denoting the average distance between quarks by $\bar{r}$,
the interaction between quarks by $\mbox{v}(m_0,n_{\mathrm{B}})$,
and assuming that each quark can only interact strongly with
other $N_0$ nearest quarks at any moment, because of
the saturation of strong interactions,
the interacting energy density $\epsilon_{\mathrm{I}}$
can be linked to density by
\begin{equation} \label{HI}
\epsilon_{\mathrm{I}}
=\frac{3}{2} N_0 n_{\mathrm{B}} \mbox{v}(m_0,\bar{r}).
\end{equation}
The average inter-quark distance
$\bar{r}$ is linked to density through
$\bar{r}=\xi /n_b^{1/3}.$
Here $\xi $ is a geometrical factor related to the way in which
we group the quarks together. In what follows, we have divided
the system into sub cubic boxes, being then $\xi=1/3^{1/3}$.
We will take $N_0=2$ since a quark has a trend to interact
strongly with other two quarks to form a baryon.
The concrete value of $N_0$ as well as the value of $\xi $
have only a marginal influence on the density
behavior of the chiral condensate.

Substituting Eq.\ (\ref{HI}) into Eqs.\ (\ref{qcmI2})
and (\ref{eImI2}), we have, respectively,
\begin{equation}  \label{qcv}
\frac{\langle\bar{q}q\rangle_{n_{\mathrm{B}}}}
     {\langle\bar{q}q\rangle_0}
=1-\frac{N_0}{2N_{\mathrm{f}}}
   \frac{n_{\mathrm{B}}}{n^*}
   \frac{\mathrm{v}}
        {m_{\mathrm{I}}},
\end{equation}
\begin{equation} \label{mvnorm}
 m F\left(\frac{k_{\mathrm{f}}}{m}\right)
  -\frac{m_0}{N_{\mathrm{f}}}
   F\left(\frac{k_{\mathrm{f}0}}{m_0}\right)
=\frac{N_0}{2N_{\mathrm{f}}}
  \mathrm{v}(m_0,n_{\mathrm{B}}).
\end{equation}

If the parameter $N_0$  diverges
faster than $k_{\mathrm{f}}$ or $n_{\mathrm{B}}^{1/3}$
at extremely higher densities,
we have
\begin{equation} \label{vlim}
\lim\limits_{n_{\mathrm{B}}\rightarrow\infty}
 \mbox{v}(m_0,\bar{r})=0.
\end{equation}
which is consistent with asymptotic freedom.

To solve Eq.\ (\ref{mIeq2}),
we need an initial condition at $m_0=m_0^*$.
 Let us suppose it to be
\begin{equation} \label{mini}
m(m_0^*,n_{\mathrm{B}})=m(n_{\mathrm{B}}).
\end{equation}
Usually, we will have
\begin{equation} \label{vini}
\mbox{v}(m_0,n_{\mathrm{B}})|_{m_0=m_0^*}=\mbox{v}(\bar{r}),
\end{equation}
where v$(\bar{r})$ is the inter-quark interaction for
the special value $m_0^*$ of the quark current mass $m_0$.

Eq.\ (\ref{mIeq2}) is difficult to solve analytically.
However, this can be done at lower densities.

Let's rewrite Eq.\ (\ref{mIeI}) as
\begin{equation} \label{mvexp}
m_{\mathrm{I}}
 =\frac{\mbox{v}(m_0,\bar{r})}
       {\frac{2}{N_0}
        f\left(\frac{k_{\mathrm{f}0}}{m_0}\right)
 +\frac{\partial\mbox{v}(m_0,\bar{r})}{\partial m_0}
          }.
\end{equation}

At lower densities, the Fermi momentum $k_{\mathrm{f}}$ is small,
so the function $F(x)$ approaches to 1. Accordingly, from
 Eq.\ (\ref{mvnorm}) we get
$
m_{\mathrm{I}}
=\frac{N_0}{2N_{\mathrm{f}}}\mbox{v}(m_0,n_{\mathrm{B}}).
$
Replacing the left hand side of Eq.\ (\ref{mvexp})
with this expression, and
integrating the resulting equation under the initial condition
given in Eq.\ (\ref{vini}), we have
\begin{eqnarray}
&& \hspace{-0.7cm}
  m_{\mathrm{I}}(m_0,n_{\mathrm{B}}) \nonumber\\
&&  \hspace{-0.7cm}
 =\frac{N_0}{2N_{\mathrm{f}}}\mbox{v}(\bar{r})
   + \int_{m_0^*}^{m_0}
 \left[1-\frac{1}{N_{\mathrm{f}}}
         f\left(\frac{k_{\mathrm{f}0}}{m_0}\right)\right] dm_0.
                     \label{mIlowexp}
\end{eqnarray}

In general, an explicit analytical solution for the condensate
 is not available, and we have to perform numerical calculations.
For a given inter-quark interaction v$(\bar{r})$,
we can first solve Eq.\ (\ref{mvnorm}) to obtain the initial condition
in Eq.\ (\ref{mini}) for the equivalent mass, then solve
the differential Eq.\ (\ref{mIeq2}), and finally calculate the quark
condensate through Eq.\ (\ref{qcv}).

There are various expressions for v($\bar{r}$) in literature,
e.g., the Cornell potential \cite{Eichten75}, the Richardson
potential \cite{Richardson79}, the so-called QCD potentials
\cite{Fischler77,Billoire80}, etc.
They are all flavor-independent.
Let's take a QCD-like interaction of the form
\begin{equation}
\mbox{v}(\bar{r})
=\sigma\bar{r}-\frac{4}{3}\frac{\alpha_s(\bar{r})}{\bar{r}}.
\end{equation}
%
 The first term $\sigma \bar{r}$ is the long-range confining part.
The second term incorporates perturbative effects. To second order
in perturbation theory, one has \cite{Fischler77,Billoire80}
\begin{equation}
\alpha_s(\bar{r})
=\frac{4\pi}{b_0\lambda(\bar{r})}
\left[
1-\frac{b_1}{b_0^2}\frac{\ln\lambda(\bar{r})}
{\lambda(\bar{r})}
 +\frac{b_2}{\lambda(\bar{r})}
\right]
\end{equation}
where \cite{Igi86}
\begin{equation}
\lambda(\bar{r})
 \equiv \ln[(\bar{r}\Lambda_{\overline{ms}})^{-2}+b]
\end{equation}
and $b_0=(11N_{\mathrm{c}}-2N_{\mathrm{f}})/3$,
$
b_1
=[34N_{\mathrm{c}}^2
 -N_{\mathrm{f}}(13N_{\mathrm{c}}^2-3)/N_{\mathrm{c}}]/3,
$
and
$
b_2
=(31N_{\mathrm{c}}
 -10N_{\mathrm{f}})/(9b_0)+2\gamma_{\mathrm{E}}
$
for SU($N_{\mathrm{c}}$) and $N_{\mathrm{f}}$ flavors.
$\gamma_{\mathrm{E}}$ is the Euler constant.

\begin{figure}[hbt]
\includegraphics[scale=0.7]{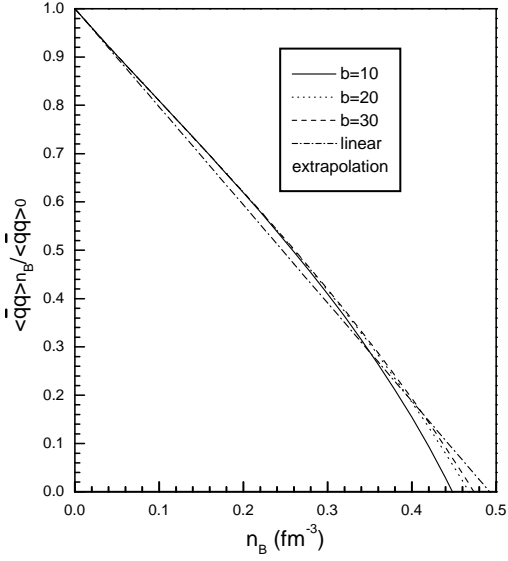}
\caption{
 Density dependence of the quark condensate in quark matter.}
\label{qcnb}
\end{figure}

Besides these constants, there are three free parameters, i.e.\
$\sigma$, $\Lambda_{\overline{ms}}$, and $b$.
The QCD scale parameter is usually taken to be
$\Lambda_{\overline{ms}}=300$ MeV.
The value for the string tension $\sigma$ from potential
models varies in the range 0.18---0.22 GeV$^2$ \cite{Veseli96},
and we take $\sigma=0.2$ GeV$^2$.
As for the parameter $b$, we take three values i.e.\ 10, 20, and 30,
in the reasonable range \cite{Igi86}.
%
%
The value of $m_0^*$ in Eq.\ (\ref{vini}) is taken to be 7.5 MeV.
The numerical results are plotted in Fig.\
\ref{qcnb}.


\medskip

In Fig.\ \ref{qcnb}, the straight line is the linear extrapolation
of Eq.\ (\ref{qcL}). It does not depend on the form of the inter-quark
interaction v$(\bar{r})$, and so is `model-independent'.
The other three lines are for $m_0=7.5$ MeV, but for different $b$
values. At lower densities, the chiral
condensate decreases linearly with increasing densities. When the
density increases, being  less than two times the nuclear saturation
density, the decreasing speed is slowed.
However, for even higher densities, it can be shown that
the condensate vanishes rapidly.


   It should be noted that if the Fermi momentum in
Eqs.\ (\ref{epsilon}) 
had not been boosted,
the main Eq.\ (\ref{qcmI2}) is still valid while
 Eqs.\ (\ref{mIeq2}) and (\ref{eImI2}), and accordingly
 Eqs.\ (\ref{mlowden}), (\ref{mvnorm}), (\ref{mvexp})--(\ref{mIlowexp}),
and especially the important equation (\ref{qcL}) and the relation
between the pion-nucleon sigma term and the current quark masses
would be different by a factor, as has been formulated in the
first part of Ref.\ \cite{ijmpa}.

\appendix
\section{Why should the effective Fermi momentum be boosted}
In this appendix, we show that the effective Fermi momentum
in the equivalent mass approach should be boosted to a higher
value.

We start from
\begin{equation}  \label{therm1}
d(VE)=Td(VS)-PdV+\mu d(Vn),
\end{equation}
which is the combination of the first and second laws of
thermodynamics. Here $n$ is the particle number density,
$E$ is the energy density, and $S$ is the entropy density.
Because the system is uniformly distributed,
the corresponding extensive quantities
are, respectively, $Vn$, $VE$, and $VS$.
$\mu$ is the chemical potential. From this expression we can get
\begin{eqnarray}
&& T=\left.\frac{dE}{dS}\right|_n,  \label{Tlab}  \\
&& P+E-TS-\mu n=-V\left.\frac{dE}{dV}\right|_{S,n}=0,
                                \label{Plab} \\
&& \mu=\left.\frac{dE}{dn}\right|_S. \label{mulab}
\end{eqnarray}

At zero temperature, the entropy becomes zero,
Eqs.\ (\ref{Plab}) and (\ref{mulab}) become, respectively,
\begin{eqnarray}
P&=&-E+\mu n, \label{Pexp} \\
dn&=&{dE}/{\mu}.   \label{dn1}
\end{eqnarray}

In our equivalent mass model, the energy density
is given with the equivalent mass as
\begin{eqnarray}
E&=&\frac{g^*}{2\pi^2}
  \int_0^{k_{\mathrm{f}}}
  \sqrt{p^2+m^2} p^2 dx  \label{E1} \\
 &=&
\frac{g^*}{16\pi^2}\left[
   k_{\mathrm{f}}
   \sqrt{k_{\mathrm{f}}^2+m^2}\
   (2k_{\mathrm{f}}^2+m^2)
   \right. \nonumber\\
 && \phantom{\frac{g}{16\pi^2}[xxxxx}\left.
  -m^4\mbox{sh}^{-1}
   \left(\frac{k_{\mathrm{f}}}{m}\right)
                 \right],       \label{E2}
\end{eqnarray}
where the Fermi momentum $k_{\mathrm{f}}$ satisfies
$k_{\mathrm{f}}\equiv\sqrt{\mu^2-m^2}$.
The degeneracy factor $g^*$ is
$N_{\mathrm{f}}$(flavor)$\times$2(spin)$\times$3(color).

>From Eq.\ (\ref{E2}) we get
$dE=\frac{\partial E}{\partial k_{\mathrm{f}}} dk_{\mathrm{f}}
   +\frac{\partial E}{\partial m}dm$.
Substituting this into Eq.\ (\ref{dn1}) then gives
\begin{eqnarray}  \label{dn2}
dn&=&\frac{g^*k_{\mathrm{f}}^2}{2\pi^2} dk_{\mathrm{f}}\nonumber\\
  &+&\frac{g^*m}{4\pi^2}
     \left[
 k_{\mathrm{f}}-\frac{m^2\mbox{sh}^{-1}(k_{\mathrm{f}}/m)}
                     {\sqrt{k_{\mathrm{f}}^2+m^2}}
    \right] dm.
\end{eqnarray}

If the mass does not depend on the density or Fermi momentum,
the second term vanishes. One then has
\begin{equation} \label{pf0}
k_{\mathrm{f}}
=\left(
\frac{6\pi^2}{g^*} n
 \right)^{1/3}.
\end{equation}
Eq. (\ref{pf0}) is the well-known expression
for the non-interacting system.
However, in the mass-density-dependent case where interactions
are treated non-pertutbatively by defining an equivalent mass,
the quark number density should be given by integrating over
both sides of Eq.\ (\ref{dn2}):
\begin{eqnarray}
n&=&\frac{g^*k_{\mathrm{f}}^3}{6\pi^2}\nonumber\\
 &+ &\frac{g^*}{4\pi^2}\int
            \left[
  k_{\mathrm{f}}-\frac{m^2\mbox{sh}^{-1}(k_{\mathrm{f}}/m)}
       {\sqrt{k_{\mathrm{f}}^2+m^2}}
            \right] mdm \label{nexp}
\end{eqnarray}
Usually the equivalent mass
is a big quantity, much larger than the current mass. Therefore,
the ratio $k_{\mathrm{f}}/m$ is small if the densities are not too
high. Let us then expand the integrand of the second
term on the right hand side of Eq.\ (\ref{nexp})
with respect to  $k_{\mathrm{f}}/m$,
taking then  the lowest order term. We get:
\begin{equation} \label{n3}
n=\frac{g^*}{6\pi^2}k_{\mathrm{f}}^3
  +\frac{g^*}{6\pi^2}\int
    \frac{k_{\mathrm{f}}^3}{m}
                   dm.
\end{equation}
Because of quark confinement and asymptotic freedom,
$m$ increases with decreasing $k_{\mathrm{f}}$.
Therefore, the simplest parametrization
should be
\begin{equation} \label{mpar}
m=m_0+\frac{C}{k_{\mathrm{f}}^Z}.
\end{equation}
with C being a constant. To be consistent with the linear confinement,
the exponent $Z$ is equal to $1$
\footnote{The parametrization to be consistent with linear
confinement is,
$m_{\mathrm{I}}\propto 1/n_{\mathrm{b}}^{1/3} \propto 1/k_{\mathrm{f}}$.
See Ref.\ \cite{Peng} (PRC61: 015201)
         }.
However, to reproduce the presently accepted value for the
pion-nucleon term (about 45 MeV), Z should be about 3/2.
Substituting Eq.\ (\ref{mpar}) into Eq.\ (\ref{n3}) then gives
\begin{equation} \label{pfI0}
k_{\mathrm{f}}
=\left(
\frac{6}{g^*}\pi^2 \frac{n}{1-Z/3}
 \right)^{1/3}.
\end{equation}

Comparing Eqs.\ (\ref{pf0}) and (\ref{pfI0}),
it is obvious that, for the same density, the
Fermi momentum of the interacting system is
different from that of the non-interacting case.
When taking $Z=3/2$,
Eq.\ (\ref{pfI0}) becomes Eq.\ (\ref{kfdef}).

\section*{Acknowledgements}
The support from the National Natural Science Foundation of China
(19905011 and 10135030), FONDECYT, Chile, (Proyectos 3010059
and 1010976), the CAS president foundation (E-26),
and the BES-BEPC fund (G6501) are gratefully acknowledged.


\begin{thebibliography}{00}




\bibitem{review}
 See, e.g.,
 G.E. Brown and M. Rho,
  Phys. Rep. {\bf 363} (2002) 85;
 M.C. Birse, J. Phys. G {\bf 20} (1994) 1537.

\bibitem{Peng}
G.X. Peng {\sl et al.}\
   Phys. Rev. C {\bf 62} (2000) 025801;
              C {\bf 61} (2000) 015201;
              C {\bf 59} (1999) 3452;
              C {\bf 56} (1997) 491.
\bibitem{Surk}
 Y. Zhang and R.K. Su,
  Phys. Rev. C {\bf 65} (2002) 035202;
  Europhys. Lett. {\bf 56} (2001) 361.

\bibitem{Ben}
 O.G. Benvenuto and G. Lugones,
    Phys.\ Rev.\ D {\bf 51} (1995) 1989;
 G. Lugones and O.G. Benvenuto,
    {\sl ibid.}\ {\bf 52} (1995) 1276.

\bibitem{Cha}
 S. Chakrabarty, S. Raha, and B. Sinha,
    Phys.\ Lett.\ B {\bf 229} (1989) 112;
 S. Chakrabarty,
    Phys.\ Rev.\ D {\bf 43} (1991) 627;
                 D {\bf 48} (1993) 1409;
                 D {\bf 54} (1996) 1306.

\bibitem{Drukarev99}
E.G. Drukarev and E.M. Levin, Nucl. Phys. A {\bf 511} (1990) 679;
E.G. Drukarev, M.G. Ryskin, V.A. Sadov\/-nikova,
 Prog. Part. Nucl. Phys. {\bf 47} (2001) 73.

\bibitem{Cohen91}
 T.D. Cohen, R.J. Furnstahl, and D.K. Griegel,
    Phys.\ Rev.\ C {\bf 45} (1992) 1881;
    Phys.\ Rev.\ Lett.\ {\bf 67} (1991) 961.

\bibitem{Gasser91}
 J. Gasser, H. Leutwyler, and M. Sainio,
    Phys.\ Lett.\ B {\bf 253} (1991) 252.


\bibitem{Gensini80}
 P.M. Gensini,
 Nuovo Cimento {\bf 60A} (1980) 221.

\bibitem{Gibbs98}
 W.R. Gibbs, Li Ai, and W.B. Kaufmann,
    Phys. Rev. C {\bf 57} (1998) 784.

\bibitem{Gasser82}
 J. Gasser, and H. Leutwyler,
    Phys. Rep. {\bf 87} (1982) 77.

\bibitem{Eichten75}
 E. Eichten, K. Gottfried, T. Kinoshita,
 J. Kogut, K.D. Lane, and T.M. Yan,
 Phys. Rev. Lett. {\bf 34} (1975) 369.

\bibitem{Richardson79}
 J.L. Richardson,
 Phys. Lett. B {\bf 82} (1979) 272.

\bibitem{Fischler77}
 W. Fischler,
 Nucl. Phys. B {\bf 129} (1977) 157.

\bibitem{Billoire80}
 A. Billiore,
 Phys. Lett. B {\bf 92} (1980) 343.

\bibitem{Igi86}
 K. Igi and S. Ono,
 Phys. Rev. D {\bf 33} (1986) 3349.

\bibitem{Veseli96}
 S. Veseli and M. Olsson,
  Phys. Lett. B {\bf 383} (1996) 109.

\bibitem{ijmpa}
 G.X.~Peng, U.~Lombardo, M.~Loewe, H.C.~Chiang, and P.Z.~Ning,
Int. J. Mod. Phys. A {\bf 18} (2003) 3151.




\end{thebibliography}
\end{document}